\documentclass[aps,prc,preprint,showpacs,floatfix]{revtex4-1}
\usepackage[dvips]{color}
\usepackage{graphicx}
\usepackage{bm}

\begin{document}

\title{Alpha decay calculations with a new formula}

\author{D.T. Akrawy$^1$ and D. N. Poenaru$^2$}
\email[]{akrawy85@gmail.com,poenaru@fias.uni-frankfurt.de}
\affiliation{$^1$Akre computer Institute, Ministry of education,
Kurdistan, Iraq, and 
Becquerel Institute for Radiation Research and
Measurements, Erbil, Kurdistan, Iraq\\and
\\$^2$Horia Hulubei National Institute of Physics and Nuclear
Engineering (IFIN-HH), \\P.O. Box MG-6, RO-077125 Bucharest-Magurele,
Romania, and Frankfurt Institute for Advanced Studies, Johann Wolfgang
Goethe University, Ruth-Moufang-Str. 1, D-60438 Frankfurt am Main, Germany}

\date{ }

\begin{abstract}

A new semi-empirical formula for calculations of $\alpha$~decay halflives is
presented.  It was derived from Royer relationship by introducing new
parameters which are fixed by fit to a set of experimental data.  We are
using three sets: set A with 130 e-e (even-even), 119 e-o (even-odd), 109
o-e, and 96 o-o, set B with 188 e-e, 147 e-o, 131 o-e, and 114 o-o, and set
C with 136 e-e, 84 e-o, 76 o-e, and 48 o-o alpha emitters.  A comparison of
results obtained with the new formula and the following well known
relationships: semFIS (semiempirical based on fission theory); ASAF
(analytical superAsymmetric fission) model, and UNIV (universal formula) is
made in terms of rms standard deviation.  We also introduced a weighted mean
value of this quantity, allowing to compare the global properties of a given
model.  For the set B the order of the four models is the following: semFIS;
UNIV; newF, and ASAF.  Nevertheless for even-even alpha emitters UNIV gives
the 2nd best result after semFIS, and for odd-even parents the 2nd is newF.
Despite its simplicity in comparison with semFIS the new formula, presented
in this article, behaves quite well, competing with the others well known
relationships

\end{abstract}


\maketitle

\section{Introduction}
\label{sec:1}

A.H. Becquerel discovered, by chance, $\alpha$, $\beta$, and $\gamma$
radioactivities --- the first experimental information coming from an atomic
nucleus.  E.  Rutherford realized that the alpha particle is $^4$He.  H. 
Geiger and J.M.  Nuttal \cite{gei11pm} gave in 1911 a simple relationship
allowing to estimate the half-lives, but only in 1928 the right explanation
was given by G.  Gamow and simultaneously by R.W.  Gurney and E.U.  Condon,
as a quantum tunnelling through the potential barrier, mostly of
electrostatic nature.

As far back as 1911, Geiger and Nuttal have found, for the members of a 
given natural radioactive family, a simple purely empirical dependence 
\cite{gei11pm} of the 
$\alpha$-decay partial half-life, $T_\alpha$, on the mean 
$\alpha$-particle range, {${\cal{R}}_\alpha$, in air (at 15$^\circ$~C
temperature and one atmosphere pressure), which may be written
as:
\begin{equation}
\log_{10} T_\alpha(s) = - 57.5 \log_{10} {\cal{R}}_\alpha (cm) + C
\end{equation} 
where $C$ depends on the series, e.g. $C=41$ for the $^{238}$U series.
One has approximately ${\cal{R}}_\alpha = 0.325 E_\alpha^{3/2}$ in which the
kinetic energy of $\alpha$~particles, $E_\alpha$, is expressed in MeV and
the range in air, ${\cal{R}}$, in cm. This relationship is now of historical
interest; the effect of atomic number, $Z$, upon decay rate is obscured.
The one-body theory of $\alpha$-decay can explain it and to a good
approximation produces a formula with an explicit dependence on the $Z$
number.  Nowadays, very often a diagram of $\log T_\alpha$ versus
$ZQ^{-1/2}$ is called Geiger-Nuttal plot \cite{boh75b}.

There are many semiempirical relationships (see for example Refs. 
\cite{fro57kd,wap59b,taa61b,vio66jinc,kel72zp,hor74np,hat90pr,brow92pr,rur82p,roy00jpg,par05app}),
allowing to estimate the disintegration period if the kinetic energy of the
emitted particle $E_\alpha = QA_d/A$ is known.  $Q$ is the released energy
and $A_d, A$ are the mass numbers of the daughter and parent nuclei. 
Alpha-decay half-life of an even-even emitter can also be easily calculated
by using the universal curves \cite{p162jpg91} or the analytical
superasymmetric (ASAF) model \cite{p195b96b}.  Some of these formulae were
only derived for a limited region of the parent proton and neutron numbers. 
Their parameters have been determined by fitting a given set of experimental
data.  Since then, the precision of the measurements was increased and new
$\alpha$-emitters have been discovered.

The description of data in the neighborhood of the magic proton and neutron
numbers, where the errors of the other relationships are large, was improved
by deriving a new formula based on the fission theory of $\alpha$-decay
\cite{p83jpl80}.  A computer program \cite{p98cpc82} allows to change
automatically the fit parameters, every time a better set of experimental
data is available.  There are many alpha emitters, particularly in the
intermediate mass region, for which both the Q-values and the half-lives are
well known \cite{wan12cpc,aud12cpc1}.  Initially it was used a set of 376
data (123 even-even (e-e), 111 even-odd (e-o), 83 odd-even (o-e), and 59
odd-odd (o-o)) on the most probable (ground state to ground state or favored
transitions) $\alpha$-decays, with a partial decay half-life
\begin{equation}
T_\alpha = (100/b_\alpha)(100/i_p)T_t
\end{equation}
where $b_\alpha$ and $i_p$, expressed in percent, represent the
branching ratio of $\alpha$-decay in competition with all other decay
modes, and the intensity of the strongest $\alpha$-transition,
respectively.

The formula given by Fr\"oman  \cite{fro57kd} is limited to the region of
even-even nuclei with $Z \geq 84$.  This formula describes well the
experimental data of nuclei with $N \geq 128$ but fails in the region of
lighter $\alpha$-emitters, which have not been available at the moment of
its derivation.  A better overall result gives a simple relationship of
Wapstra et al.  \cite{wap59b} also valid for even-even nuclei with $Z \geq
85$.  In the new variant derived by A.  Brown \cite{brow92pr} for nuclei
with $Z\geq 72$ the agreement with experimental data is not bad for nuclei
with $Z\geq 72$, but large errors are obtained for lighter parent nuclei.

The formula presented by Taagepera and Nurmia  \cite{taa61b} remains one of
the best.  It is exceeded by a variant due to Keller and M\"unzel
\cite{kel72zp}.  Viola and Seaborg \cite{vio66jinc} introduced a
relationship which gives excellent agreement in the region of actinides but
it underestimates the lifetimes of lighter nuclei, in contrast with
overestimations obtained with the first of the above mentioned formulae.

In the region of superheavy nuclei the majority of researchers prefer to use
Viola-Seaborg formula.  Very recently for nuclei with $Z=84-110$ and
$N=128-160$, for which both $Q_\alpha ^{exp}$ and $T_{exp}$ experimental
values are available, new optimum parameter values \cite{par05app} have been
determined.  The average hindrance factors for 45 o-e ($Z=85-107$), 55 e-o
($Z=84-110$), and 40 o-o ($Z=85-111$, $N=129-161$) nuclei were determined to
be $C_V^p=0.437$, $C_V^n=0.641$, and $C_V^{pn}=1.024$.  In this way
$T_{exp}$ were reproduced by the Viola-Seaborg formula within a factor of
1.4 foe e-e, 2.3 for o-e, 3.7 for e-o and 4.7 for o-o nuclei, respectively. 
Good results were obtained with a formula due to Royer \cite{roy00jpg}
having 12 parameters $a,b,c$ for e-e, e-o, o-e and o-o nuclei.  Shell
effects were not taken into account; nuclei with neutron number close to the
shell closures $N=152$ and $162$ (namely 3 nuclei with $N=151 (Z=96,
98,100)$ one with $N=153 \ Z=98$, and one with $N=161 \ Z=110$) have been
omitted in the fitting procedure.  Other omission of 3 o-e nuclei with
$Z=97, N=146,148$ and $Z=101, N=154$ was motivated by a large deviation from
the average behaviour.  A simple version of the Viola-Seaborg formula was
proposed by Parkhomenko and Sobiczewski \cite{par05app}.

Since 1979 one of us (DNP) considered $\alpha$~decay a superasymmetric
fission process \cite{p76jpg79,p78jpl79}.  Consequently a new semiempirical
formula for the alpha decay halflives \cite{p83jpl80} was a straightforward
finding \cite{p83jpl80,p101jp83}.  Moreover, the analytical and numerical
superasymmetric fission (ASAF \cite{p103jp84} and NUSAF) models were used
together with fragmentation theory developed by the Frankfurt School, and
with penetrability calculations like for $\alpha$~decay, to predict cluster
(or heavy particle) radioactivity
\cite{ps84sjpn80,enc95,p109jpg84,p115pr85,ps117prl85,p309prl11}.  The
extended calculations \cite{p123adnd86,p160adnd91,p268pr06} have been used
to guide the experiments and as a reference for many theoretical
developments.  A series of books and chapters in books, e.g. 
\cite{p140b89,p172b96,p195b96,p193b97,p302bb10} are also available.  A
computer program \cite{p98cpc82} gives us the possibility to improve the
parameters of the ASAF model in agreement with a given set of experimental
data.  The UNIV (universal curve) model was updated in 2011
\cite{p308prc11}.

The interest for $\alpha$D is strongly simulated by the search for heavier
and heavier superheavies (SHs) --- nuclides with $Z>103$, produced by fusion
reactions \cite{khu14prl,ham13arnps,due10prl,mor07jpsjb} who may be
identified easily if a chain of $\alpha$D leading to a known nucleus may be
measured.  Recently it was shown that for superheavy nuclei with atomic
numbers $Z>121$ \cite{p309prl11,p315prc12} $\alpha$D may be stronger than CD
or spontaneous fission.

A very interesting result was reported by Y.Z.  Wang et al. \cite{wan15prc},
who compared 18 such formulae in the region of superheavy nuclei.  They
found: ``SemFIS2 formula is the best one to predict the alpha-decay
half-lives ...  In addition, the UNIV2 formula with fewest parameters and
the VSS, SP and NRDX formulas with fewer parameters work well in prediction
on the SHN alpha-decay half-lives
\cite{p274el07,vio66jinc,sob07ppnp,par05app,ni08prc}.''

In this work we intend to study how may be improved a formula developed by
G.  Royer \cite{roy10np} by adding few parameters fitted to experimental
data.  We shall use three data sets, say: A (130 e-e, 119 e-o, 109 o-e, and
96 o-o), set B (188 e-e, 147 e-o, 131 o-e, and 114 o-o), and set C with 136
e-e, 84 e-o, 76 o-e, and 48 o-o alpha emitters.set C with 136 e-e, 84 e-o,
76 o-e, and 48 o-o alpha emitters.  The set A was developed by one of us
(DA), the set B belongs to DNP's group, and the set C was taken from G. 
Royer \cite{roy10np}; few Q-values have been updated using the AME12
evaluation of experimental atomic masses \cite{wan12cpc}.  Comparison with
ASAF, UNIV, and semFIS will be made using both A, B and C data sets.

\section{New formula}

\begin{figure}[ht]
\begin{center}
\includegraphics[width=15cm]{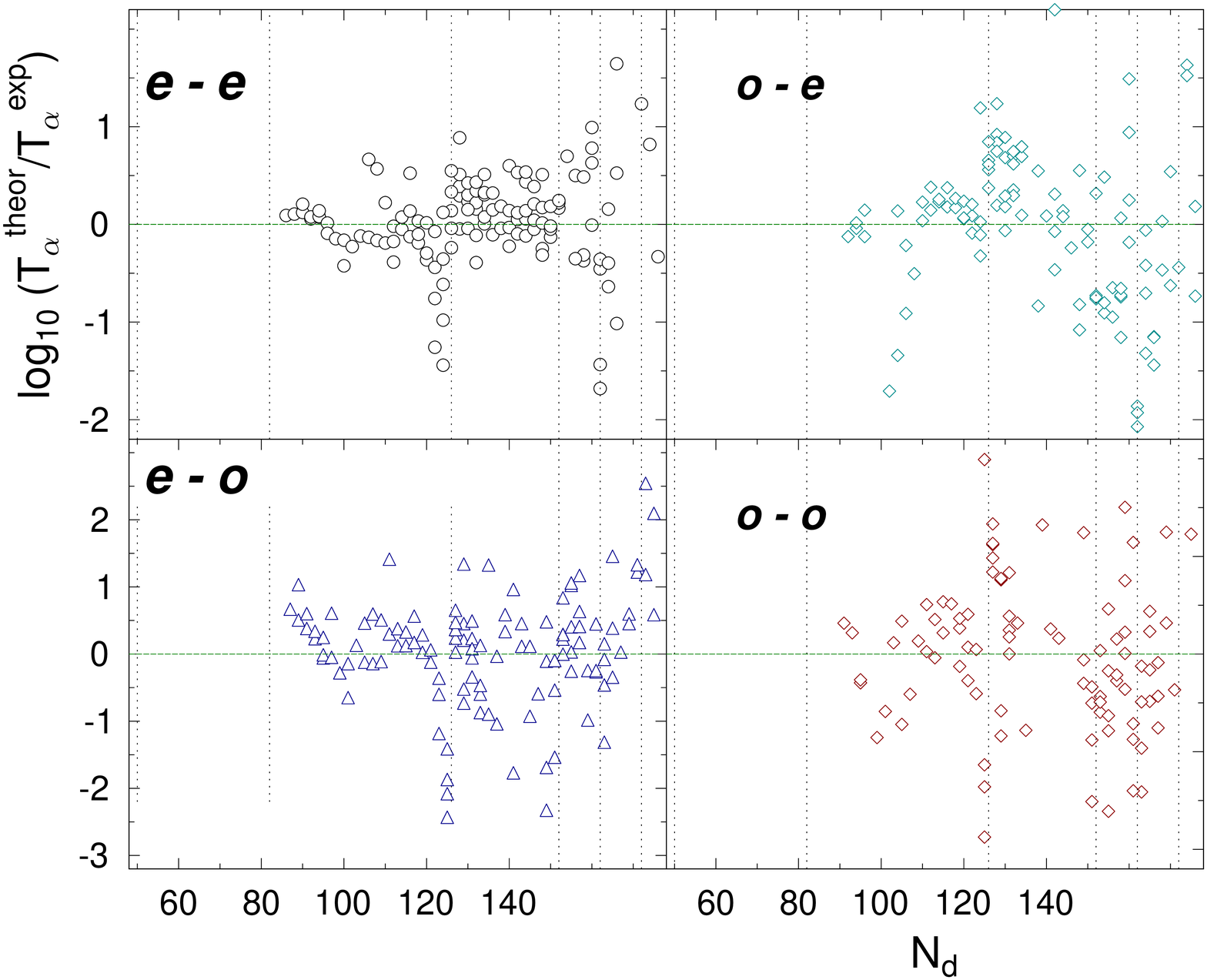} 
\end{center}
\caption{The differences of $\log_{10}T_{theor} - \log_{10}T_{exp}$ in four
groups of alpha emitters versus the neutron number of the daughter $N_d$. 
New formula.  Vertical dashed lines are marking magic numbers of neutrons,
either spherical or deformed.
\label{difakr}}
\end{figure}

The Royer formula \cite{roy10np} is defined as
\begin{equation}
T_{1/2} = a + bA^{1/6}\sqrt{Z} + \frac{cZ}{\sqrt{Q_\alpha}}
\end{equation}
with initial parameters $a=-27.657; -28.408; -27.408, and -24.763$,
$b=-0.966; -0.920;$ $-1.038, and -0.907$, and $c=1.522; 1.519; 1.581, and
1.410$ for e-e, e-o, o-e, and o-o, respectively. The rms standard deviation
for 130 e-e, 119 e-o, 109 o-e, and 96 o-o was $\sigma = 0.560, 1.050,
0.871, $ and $0.926$, respectively.

The new relationship is obtained by introducing $I=(N-Z)/A$ and the
new parameters d and e:
\begin{equation}
T_{1/2} = a + bA^{1/6}\sqrt{Z} + \frac{cZ}{\sqrt{Q_\alpha}} + dI + eI^2
\end{equation}
where initially for the set A the parameters a, b, c, d, e are given in
table \ref{tabca}.

Before optimization, with our set of 580 $\alpha$~emitters, and the initial
values of the parameters $a=-27.989, b=-0.940, c=1.532, d=-5.747, e=11.336 $
for even-even nuclei we got the following values of rms standard deviations,
$\sigma = 0.5547$.  After optimization for e-e emitters, with
$a=-27.837,b=-0.94199975, c=1.5343, d=-5.7004, e=8.785$ the agreement was
improved: $\sigma = 0.540$.  The order of optimization of the 5 parameters
was: a; e; d; c, and b.  

We may compare the results obtained by using the set A, B and C in tables
\ref{tabca} and \ref{tabcb}.  
We can see a slight improvement by using the set B.

\begin{table}[h] 
\caption{Optimization of coefficients using the set A (454 data). 
New formula. 
\label{tabca}}       
\begin{center}
\begin{ruledtabular}
\begin{tabular}{cccccccc}
Group&n&$\sigma$&a&b&c&d&e\\  \hline
e-e & 130 &0.557&-27.884&-0.952&1.533&~-4.101&~~6.285 \\ 
e-o & 119 &0.961&-26.160&~-1.140&1.559&~17.756&~37.055\\
o-e & 109 &0.816&-27.800&~-0.897&~1.535&~15.319&~30.443 \\
o-o & ~96 &0.915&-24.292&~-0.911&~1.409&~-3.418&~~7.640\\  
\end{tabular}
\end{ruledtabular}
\end{center}
\end{table} 

\begin{table}[h] 
\caption{Optimization of coefficients using the set B.
\label{tabcb}}       
\begin{center}
\begin{ruledtabular}
\begin{tabular}{cccccccc}
Group&n&$\sigma$&a&b&c&d&e\\  \hline
e-e & 188 &0.540   &-27.837&-0.94199975&1.5343 &-5.7004 &8.785 \\ 
e-o & 147 &0.678& -28.2245 & -0.8629   &1.53774&-21.145 &53.890 \\
o-e & 131 &0.522&  -26.8005& -1.10783  & 1.5585&14.8525 &-30.523 \\
o-o & 114 &0.840&  -23.6354& -0.891    & 1.404 &-12.4255& 36.9005\\  
\end{tabular}
\end{ruledtabular}
\end{center}
\end{table} 

\begin{table}[h] 
\caption{Optimization of coefficients using the set C.
\label{tabc}}       
\begin{center}
\begin{ruledtabular}
\begin{tabular}{cccccccc}
Group&n&$\sigma$&a&b&c&d&e\\  \hline
e-e & 136 &0.298&-26.661&-1.151&1.591&~14.680&-50.779 \\ 
e-o & ~84 &0.914&-32.567&-0.851&1.691&-32.307&111.787 \\
o-e & ~76 &0.859&-30.303&-1.006&1.749&-41.203&124.968 \\
o-o & ~48 &0.810&-25.542&-1.139&1.619&-21.016&109.613 \\  
\end{tabular}
\end{ruledtabular}
\end{center}
\end{table} 

\section{ASAF}

The half-life of a parent nucleus $AZ$ against the split into a cluster $A_e
Z_e$ and a daughter $A_d Z_d$
\begin{equation}
T = [(h \ln 2)/(2E_{v})] exp(K_{ov} + K_{s})
\end{equation}
is calculated by using the WKB quasiclassical approximation, according to
which the action integral is expressed as
\begin{equation}
K =\frac{2}{\hbar}\int_{R_a}^{R_b}\sqrt{2B(R)E(R)} dR
\end{equation}
with $B=\mu$ --- the reduced mass, $K=K_{ov}+K_s$, and $E(R)$ replaced by
$[E(R)-E_{corr}] - Q$.  $E_{corr}$ is a correction energy similar to the
Strutinsky \cite{str67np} shell correction, also taking into account the
fact that Myers-Swiatecki's liquid drop model (LDM) \cite{mye66np}
overestimates fission barrier heights, and the effective inertia in the
overlapping region is different from the reduced mass.  The turning points
of the WKB integral are:
\begin{equation}
R_a = R_i + (R_t - R_i)[(E_v + E^*)/E_b^0]^{1/2}
\end{equation}
\begin{equation}
R_b = R_tE_c \{ 1/2+[1/4+(Q+E_v+E^*)E_l/E_c^2]^{1/2} \} /(Q+E_v+E^*)
\end{equation}
where $E^*$ is the excitation energy concentrated in the separation degree
of freedom, $R_i=R_0-R_e$ is the initial separation distance, $R_t=R_e+R_d$
is the touching point separation distance, $R_j=r_0A_j^{1/3}$ $(j=0, e, d ;
\; r_0=1.2249$~fm) are the radii of parent, emitted and daughter nuclei, and
$E_b^0=E_i-Q$ is the barrier height before correction.  The interaction
energy at the top of the barrier, in the presence of a non negligible angular
momentum, $l\hbar$, is given by:
\begin{equation}
E_i=E_c+E_l=e^2Z_eZ_d/R_t+\hbar^2l(l+1)/(2\mu R_t^2)
\end{equation}
The two terms of the action integral $K$, corresponding to the overlapping
($K_{ov}$) and separated ($K_s$) fragments, are calculated by analytical  
formulas (approximated for $K_{ov}$ and exact for $K_s$ in case of separated
spherical shapes within the LDM):
\begin{equation}
K_{ov}=0.2196(E^0_bA_eA_d/A)^{1/2}(R_t-R_i)\left [  \sqrt{1-b^2}-
b^2 \ln \frac{1+\sqrt{1-b^2}}{b} \right ]
\end{equation}                           
\begin{equation}                         
K_{s}=0.4392[(Q+E_v+E^*)A_eA_d/A]^{1/2}R_b J_{rc} \;   ;  \;
b^2=(E_v+E^*)/E_b^0
\end{equation} 
\begin{eqnarray}
J_{rc}  & = & (c) \arccos \sqrt{(1-c+r)/(2-c)} - [(1-r)(1-c+r)]^{1/2}
                \nonumber \\
        & + & \sqrt{1-c} \ln \left [
  \frac{2\sqrt{(1-c)(1-r)(1-c+r)} +2-2c+cr}{r(2-c)} \right ]
\end{eqnarray}
where $r=R_t/R_b$ and $c=rE_c/(Q+E_v+E^*)$. In the absence of the
centrifugal contribution ($l=0$), one has $c=1$.

The choice $E_{v} = E_{corr}$ allows to get a smaller number of parameters. 
It is evident that, owing to the exponential dependence, any small variation
of $E_{corr}$ induces a large change of T, and thus plays a more important
role compared to the preexponential factor variation due to $E_{v}$.  Shell
and pairing effects are included in $E_{corr}=a_i(A_e)Q$ ($i=1,2,3,4$ for
even-even, odd-even, even-odd, and odd-odd parent nuclei).  For a given
cluster radioactivity there are four values of the coefficients $a_i$, the
largest for even-even parent and the smallest for the odd-odd one (see
figure~1 of \cite{p160adnd91}).  The shell effects for every cluster
radioactivity is implicitly contained in the correction energy due to its
proportionality with the Q value, which is maximum when the daughter nucleus
has a magic number of neutrons and protons.

With only few exceptions, in the region of nuclei far from stability,
measured $\alpha$-decay partial half-lives are not available.  In principle
we can use the ASAF model to estimate these quantities.  Nevertheless,
slightly better results can be obtained by using semFIS \cite{p195b96a}. 
The potential barrier shape similar to that we considered within the ASAF
model was calculated by using the macroscopic-microscopic method
\cite{p266pr06}, as a cut through the PES at a given mass asymmetry, usually
the $^{208}$Pb valley or not far from it.

Before any other model was published, there were estimations of the
half-lives for more than 150 decay modes, including all cases experimentally
confirmed until now.  A comprehensive table was produced by performing
calculations within that model.  Subsequently, the numerical predictions of
the ASAF model have been improved by taking better account of the pairing
effect in the correction energy, deduced from systematics in four groups of
parent nuclei (even - even, odd - even, even - odd and odd - odd).  In a new
table, published in 1986, cold fission as cluster emission has been
included.  The systematics was extended in the region of heavier emitted
clusters (mass numbers A$_{e} >$ 24), and of parent nuclei far from
stability and superheavies.  Since 1984, the ASAF model results have been
used to guide the experiments and to stimulate other theoretical works.
\begin{table}[h] 
\caption{Parameters of ASAF model; data --- the set A.
\label{asafa}}       
\begin{center}
\begin{ruledtabular}
\begin{tabular}{cccccccc}
Group&n&C$_{i1}$&C$_{i2}$&C$_{i3}$&C$_{i4}$&D$_i$&y50$_i$\\  \hline
e-e & 130 &0.985286&0.0179960&0.027056&0.030373&0.001215&-0.018261 \\ 
e-o & 119 &1.011020&-0.027134&0.074588&0.051785&0.141828&-0.139752 \\
o-e & 109 &0.990173&~0.063476&0.112658&0.041385&0.087355&-0.168133 \\
o-o & ~96 &0.989577&~0.025327&0.226867&0.043133&0.989809&-0.296017 \\  
\end{tabular}
\end{ruledtabular}
\end{center}
\end{table} 
\begin{table}[h] 
\caption{Parameters of ASAF model for the two other data sets: B, and C.
\label{asafp}}       
\begin{center}
\begin{ruledtabular}
\begin{tabular}{cccccccc}
Group&n&C$_{i1}$&C$_{i2}$&C$_{i3}$&C$_{i4}$&D$_i$&y50$_i$\\  \hline
e-e & 188 or 136 &5.62810&-2.81718&1.53065&-3.97164&5.666667&0.03680 \\ 
e-o & 147 or 84 &4.28000&-2.16122&1.55363&-4.07848&3.933333&0.03200 \\
o-e & 131 or 76 &4.85400&-2.63110&1.56753&-2.92537&4.800000&0.03440 \\
o-o & 114 or 48 &3.70000&-1.66474&1.46448&-4.60082&2.933333&0.03000 \\  
\end{tabular}
\end{ruledtabular}
\end{center}
\end{table} 

\begin{table}[h] 
\caption{Standard deviations for ASAF model with the set A, B, and C.
\label{asafcs}}       
\begin{center}
\begin{ruledtabular}
\begin{tabular}{cccccccc}
Group&n&$\sigma_{ASAF}^A$ &n&$\sigma_{ASAF}^B$&n&$\sigma_{ASAF}^C$ \\  \hline
e-e &130&0.731& 188& 0.415&136&0.438 \\ 
e-o &119&1.069& 147& 0.713& ~84&1.426 \\
o-e &109&1.044& 131& 0.637& ~76&1.336 \\
o-o &~96&1.041& 114& 0.876& ~48&1.069 \\  
\end{tabular}
\end{ruledtabular}
\end{center}
\end{table} 

\section{UNIV (Universal Formula)}

In cluster radioactivity and $\alpha$-decay the (measurable) decay constant
$\lambda = \ln 2 /T$, can be expressed as a product of three (model
dependent) quantities
\begin{equation}   
\lambda = \nu S P_s
\end{equation}
where $\nu $ is the frequency of assaults on the barrier per second, $S$ is
the preformation probability of the cluster at the nuclear surface, and
$P_s$ is the quantum penetrability of the external potential barrier.  The
frequency $\nu$ remains practically constant, the preformation differs from
one decay mode to another but it is not changed very much for a given
radioactivity, while the general trend of penetrability follows closely that
of the half-life.  The external part of the barrier (for separated
fragments), essentially of Coulomb nature, is much wider than the internal
one (still overlapping fragments).

According to Ref.  \cite{p167ps91} the preformation probability can be
calculated within a fission model as a penetrability of the internal part of
the barrier, which corresponds to still overlapping fragments.  One may
assume as a first approximation, that preformation probability only depends
on the mass number of the emitted cluster, $S=S(A_e)$.  The next assumption
is that $\nu(A_{e}, Z_{e}, A_{d}, Z_{d}) = $~constant.  In this way one
arrives at a single straight line {\em universal curve} on a double
logarithmic scale
\begin{equation}
\log T = -\log P_s - 22.169 + 0.598(A_{e}-1)
\end{equation}
where
\begin{equation} 
-\log P_s = c_{AZ} \left [\arccos \sqrt{r} - \sqrt{r(1-r)}\right ]
\end{equation}
with $c_{AZ}= 0.22873({\mu}_{A}Z_{d}Z_{e}R_{b})^{1/2}$,
$r=R_{t}/R_{b}$, $R_{t}=1.2249(A_{d}^{1/3}+A_{e}^{1/3})$,
$R_{b}=1.43998Z_{d}Z_{e}/Q$, and $\mu_A = A_d A_e / A$.

Sometimes this universal curve is misinterpreted as being a Geiger-Nuttal
plot.  Nowadays by Geiger-Nuttal diagram one understands a plot of $\log T$
versus $ZQ^{-1/2}$, or versus $Q^{-1/2}$.  In this kind of systematics the
experimental points are scattered.  Nevertheless, for a given atomic number,
$Z$, or for the members of a natural radioactive series, it is still
possible to get a single straight line.

The strong shell effect at the magic neutron number $N=126$, which was
ignored when the approximation $S = S(A_e)$ was made to give a pronounced
underestimation of the half-lives in the neighborhood of $N=126$.

For $\alpha$-decay of even-even nuclei, $A_e=4$, one has
\begin{equation}
\log T = -\log P_s + c_{ee}
\end{equation}
where $c_{ee}=\log S_{\alpha} - \log \nu + \log (\ln 2) = -20.375$.  We can
find new values for $c_{ee}$ and we also can extend the relationship to
even-odd, odd-even, and odd-odd nuclei, by fitting a given set of
experimentally determined alpha decay data.

\section{semFIS (Semiempirical relationship based on fission theory of
$\alpha$-decay)}

Mainly the $Z$ dependence was stressed by all formulae, in spite of
strong influence of the neutron shell effects. The neighborhood of
the magic numbers of nucleons is badly described by all these
relationships.
\begin{table}[h] 
\caption{Standard deviations of calculated half-lives ($\log_{10}T_\alpha
(s)$) with UNIV after optimization compared to experimental data in four
groups of parent nuclei: even-even; even-odd; odd-even, and odd-odd. The set
A, B, and C.
\label{tabsemf}} 
\begin{center}
\begin{ruledtabular}
\begin{tabular}{ccccccc}
Group& n&$\sigma_{UNIV}^A$& n&$\sigma_{UNIV}^B$& n&$\sigma_{UNIV}^C$ \\
\hline
e-e &130&0.560            &188&0.223           &136&0.287 \\
e-o &119&1.050&147&0.533&~84&1.384 \\
o-e &109&0.871&131&0.442&~76&1.269 \\
o-o &~96&0.926&114&0.609&~48&1.494 \\ 
\end{tabular}
\end{ruledtabular}
\end{center}
\end{table}

The SemFIS formula  based on the fission theory of $\alpha$-decay gives
\begin{equation}
\log T = 0.43429K_s\chi - 20.446
\label{eq:pf}
\end{equation}
where 
\begin{eqnarray}
K_s & = & 2.52956 Z_{da}[A_{da}/(AQ_{\alpha})]^{1/2}[\arccos \sqrt{x} 
- \sqrt{x(1-x)}] \; ; \nonumber \\
   & x & =0.423Q_{\alpha}(1.5874+A_{da}^{1/3})/Z_{da} 
\end{eqnarray}
\begin{table}[pt]
\caption{B$_k$ parameters of semFIS formula obtained by fitting the data
evaluated by Rytz.}
\begin{tabular}{ll|rrrrrr} \hline
Group&$\sigma$ & B$_1$   &  B$_2$   &  B$_3$   &  B$_4$   &  B$_5$   &B$_6$ \\ \hline
e-e&0.223&0.993119 &-0.004670 &0.017010 &0.045030 & 0.018102&-0.025097\\
o-e&0.533&1.000560 & 0.010783 &0.050671 &0.013919 & 0.043657&-0.079999\\
e-o&0.442&1.017560 &-0.113054 &0.019057 &0.147320 & 0.230300&-0.101528\\
o-o&0.609&1.004470 &-0.160560 &0.264857 &0.212332 & 0.292664&-0.401158\\
\hline
\end{tabular}
\label{tab1}
\end{table}
\begin{table}[pt]
\caption{B$_k$ parameters of semFIS formula obtained by fitting the set C.}
\begin{tabular}{ll|rrrrrr} \hline
Group&$\sigma$ & B$_1$   &  B$_2$   &  B$_3$   &  B$_4$   &  B$_5$   &B$_6$ \\ \hline
e-e&0.287&0.993119 &-0.004670 &0.017010 &0.045030 & 0.018102&-0.025097\\
o-e&1.384&1.017560 &-0.113054 &0.019057 &0.147320 & 0.230300&-0.101528\\
e-o&1.269&1.000560 &~0.010783 &0.050671 &0.013919 & 0.043657&-0.079999\\
o-o&1.494&1.004470 &-0.160560 &0.264857 &0.212332 & 0.292664&-0.401158\\
\hline
\end{tabular}
\label{tabbc}
\end{table}
and the numerical coefficient $\chi$, close to unity, is a 
second-order polynomial
\begin{equation}
\chi=B_1+B_2y+B_3z+B_4y^2+B_5yz+B_6z^2 
\end{equation}
in the reduced variables $y$ and $z$, expressing the distance from the
closest magic-plus-one neutron and proton numbers $N_i$ and $Z_i$:
\begin{equation}
y \equiv (N-N_i)/(N_{i+1} - N_i) \; ; \; N_i < N \le N_{i+1}
\end{equation}
\begin{equation}
z \equiv (Z-Z_i)/(Z_{i+1} - Z_i) \; ; \; Z_i < Z \le Z_{i+1}
\label{eq:pl}
\end{equation}
with $N_i=...., 51, 83, 127, 185, 229, .....$ , $Z_i=...., 29, 51, 83, 115,
.....$ , and $Z_{da}=Z-2$ , $A_{da}=A-4$ .  The coefficients $B_i$ obtained
by using a high-quality selected set of alpha-decay data \cite{ryt91adnd}
are given in the Table~\ref{tab1}.  Better agreement with experimental
results are obtained in the region of superheavy nuclei by introducing other
values of the magic numbers plus one unit for protons (suggesting that the
next magic number of protons could be 126 instead of 114): $Z_i=...., 83,
127, 165, .....$

With the SemFIS formula, Rurarz  \cite{rur82p} have made predictions for
nuclei far from stability with $62 <Z < 76$.  In the variant of Ref. 
\cite{hat90pr} the shell effect on the formation factor was approximated by
an empirical relationship.

Practically for even-even nuclei, the increased errors in the neighborhood
of $N=126$, present in all other cases, are smoothed out by SemFIS~formula
using the second order polynomial approximation for $\chi$.  They are still
present for the strongest $\alpha$-decays of some even-odd and odd-odd
parent nuclides.  In fact for non-even number of nucleons the structure
effects became very important, and they should be carefully taken into
account for every nucleus, not only globally.  An overall estimation of the
accuracy, gives the standard rms deviation of $\log T$ values:
\begin{equation}
\sigma = \left\{\sum_{i=1}^n[\log
(T_i/T_{exp})]^2/(n-1)\right\}^{1/2}
\end{equation}
The parameters $\{B_k\}$ of the SemFIS~formula could be automatically
improved, for a given set of experimental data, by using the computer
program described in the Ref.  \cite{p98cpc82}.  The partial $\alpha$-decay
half-lives plotted in this figure are lying in the range of $10^{-7}$ to
$10^{25}$ seconds.  One can see the effect of the spherical and deformed
neutron magic numbers of the daughter nuclei $N_d=126, 152, 162$
particularly clear for even-even and even-odd nuclides.  For the large set
of alpha emitters the following values of the rms errors have been obtained:
$\log T$: 0.19 for SemFIS~formula; 0.33 for the universal curve; 0.39 for
ASAF model, and 0.43 for numerical superasymmetric (NuSAF) model
\cite{p195b96b}.
\begin{figure}[ht]
\begin{center}
\includegraphics[width=15cm]{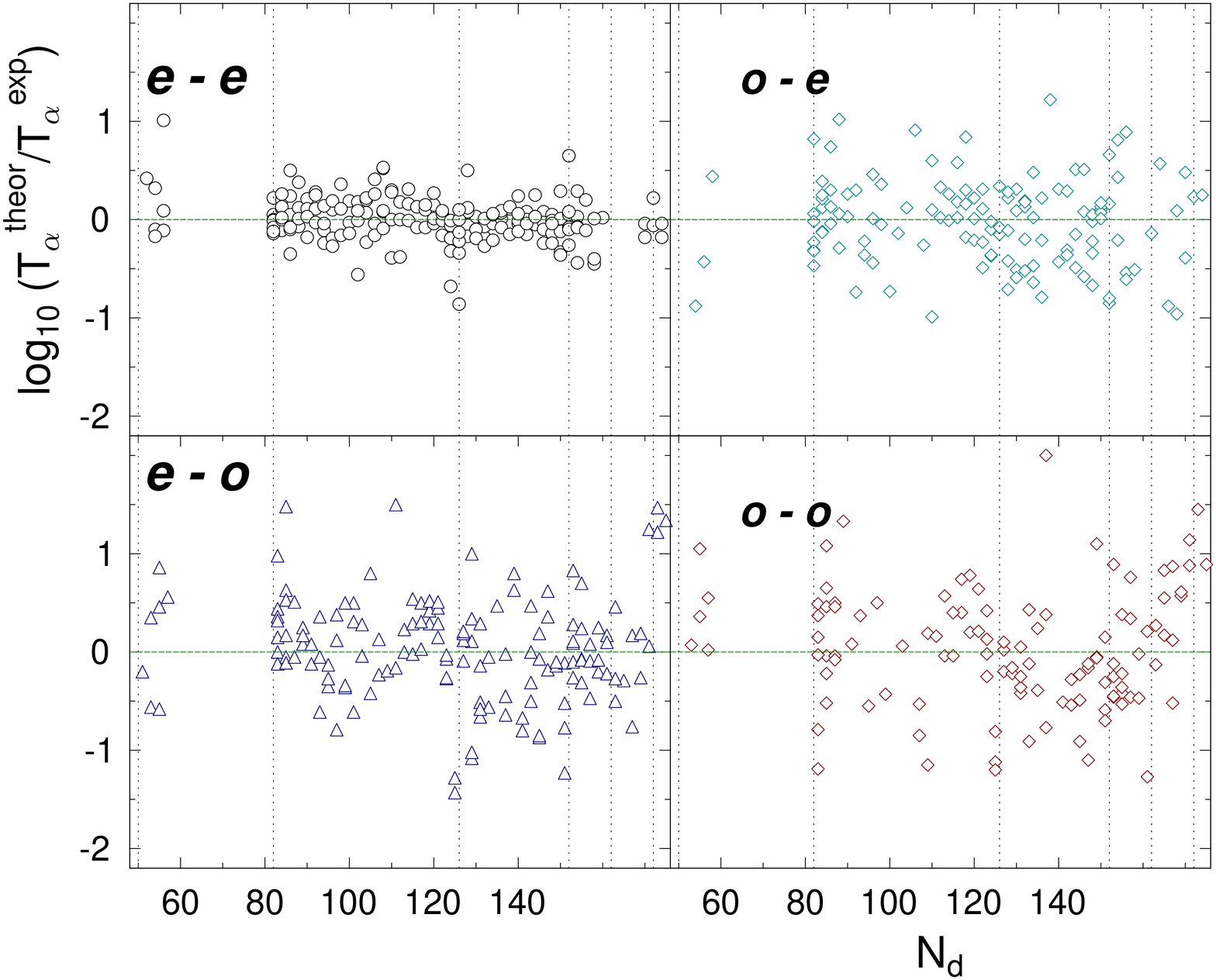} 
\end{center}
\caption{The differences of $\log_{10}T_{theor} - \log_{10}T_{exp}$ in four
groups of alpha emitters versus the neutron number of the daughter $N_d$. 
semFIS formula.  Vertical dashed lines are marking magic numbers of
neutrons, either spherical or deformed.
\label{difsem}}
\end{figure}

There are many parameters of the SemFIS formula introduced in order to
reproduce the experimental behaviour around the magic numbers of protons and
neutrons, which could be a drawback in the region of light and intermediate
alpha emitters.  In the region of superheavies these characteristics may be
conveniently used to get informations concerning the next magic numbers of
protons and neutrons which are not well known until now.  When accurate
experimental values of $Q$ and $T$ are available in the region centered on
$Z=114-126$, $N=172-184$, the SemFIS formula may be used to estimate whether
the right value of the spherical magic number is $Z=114$, $Z=120$, $Z=126$,
and $N=172$, or $N=184$, due to the high sensitivity of $\chi$ to the values
of $Z_i$ and $N_i$ (see eqs.~\ref{eq:pf}--\ref{eq:pl}).

In figure \ref{difsem} we plotted the individual errors: differences of
$\log_{10}T_{theor} - \log_{10}T_{exp}$ in four groups of alpha emitters. 
Vertical dashed lines are marking magic numbers of neutrons, either
spherical (50, 82, 126) or deformed (152, 162, 172). One can see that the
best result is always obtained for even-even alpha emitters.

Superheavy (SH) nuclei, with atomic numbers $Z=104-118$, are decaying mainly
by $\alpha $~decay and spontaneous fission.  They have been produced in cold
fusion or hot fusion ($^{48}$Ca projectile) reactions
\cite{khu14prl,ham13arnps,due10prl,mor07jpsjb}.  In a systematic study of
$\alpha $-decay energies and half-lives of superheavy nuclei it was shown
\cite{wan15prc} that our semFIS (semiempirical formula based on fission
theory) and UNIV (universal curve) are the best among 18 calculations
methods of $\alpha $~decay half-lives.  For some isotopes of even heavier
SHs, with $Z>121$, there is a good chance for cluster decay modes to compete
\cite{p315prc12,p309prl11}.

\section{Comparison of results obtained with the new formula, semFIS, UNIV,
and ASAF.}

\begin{figure}[ht]
\begin{center}
\includegraphics[width=15cm]{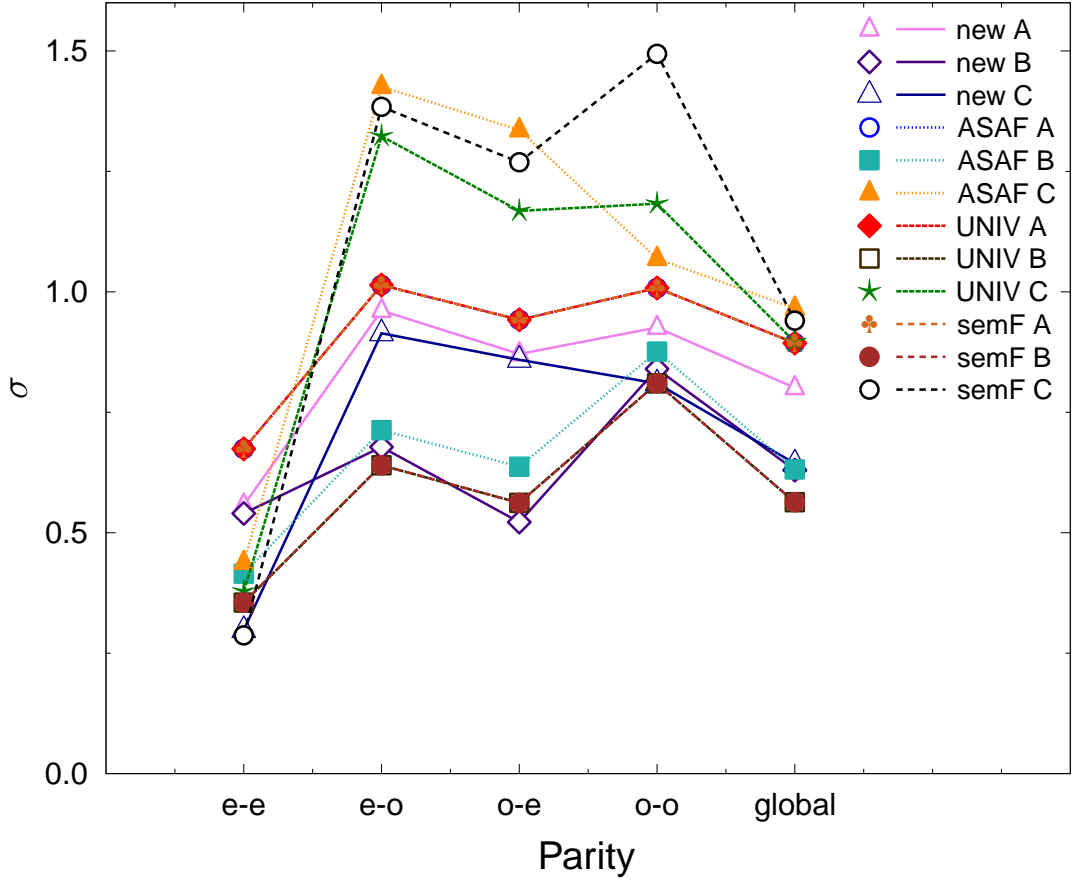} 
\end{center}
\caption{Standard rms deviations of the four models: new=newF; ASAF; UNIV,
and semFIS in four groups of e-e, e-o, o-e, o-o emitters, as well as the
global parameter, uding the three sets of experimental data: A; B, and C.
\label{stand}}
\end{figure}
We present in figure \ref{stand} the results obtained using
the sets A, B, and C, respectively. A global indicator for a given model
could be the weighted mean value
\begin{equation}
\sigma_{newF}^A=\frac{130\sigma_{e-e} + 119\sigma_{e-o} + 109\sigma_{o-e} + 
96\sigma_{o-o}}{454} = 0.8008 
\end{equation}

\begin{equation}
\sigma_{newF}^B=\frac{188\sigma_{e-e} + 147\sigma_{e-o} + 131\sigma_{o-e} +   
114\sigma_{o-o}}{580} =0.6299
\end{equation}

\begin{equation}
\sigma_{newF}^C=\frac{136\sigma_{e-e} + 84\sigma_{e-o} + 76\sigma_{o-e} +   
48\sigma_{o-o}}{344} = 0.6438 
\end{equation}

Similarly for the other models
\begin{equation}
\sigma_{ASAF}^A=\frac{130\sigma_{e-e} + 119\sigma_{e-o} + 109\sigma_{o-e} + 
96\sigma_{o-o}}{457} = 0.9591 
\end{equation}

\begin{equation}
\sigma_{ASAF}^B=\frac{188\sigma_{e-e} + 147\sigma_{e-o} + 109\sigma_{o-e} +   
114\sigma_{o-o}}{580} = 0.6313
\end{equation}

\begin{equation}
\sigma_{ASAF}^C=\frac{136\sigma_{e-e} + 84\sigma_{e-o} + 76\sigma_{o-e} +   
48\sigma_{o-o}}{344} = 0.9657 
\end{equation}

\begin{equation}
\sigma_{UNIV}^A=\frac{130\sigma_{e-e} + 119\sigma_{e-o} + 109\sigma_{o-e} + 
96\sigma_{o-o}}{457} = 0.8937 
\end{equation}

\begin{equation}
\sigma_{UNIV}^B=\frac{188\sigma_{e-e} + 147\sigma_{e-o} + 131\sigma_{o-e} +   
114\sigma_{o-o}}{580} = 0.5634
\end{equation}

\begin{equation}
\sigma_{UNIV}^C=\frac{136\sigma_{e-e} + 84\sigma_{e-o} + 76\sigma_{o-e} +   
48\sigma_{o-o}}{344} = 0.8958
\end{equation}

\begin{equation}
\sigma_{semFIS}^A=\frac{130\sigma_{e-e} + 119\sigma_{e-o} + 109\sigma_{o-e} + 
96\sigma_{o-o}}{457} = 0.9135
\end{equation}

\begin{equation}
\sigma_{semFIS}^B=\frac{188\sigma_{e-e} + 147\sigma_{e-o} + 131\sigma_{o-e} +   
114\sigma_{o-o}}{580} =  0.4269
\end{equation}

\begin{equation}
\sigma_{semFIS}^C=\frac{136\sigma_{e-e} + 84\sigma_{e-o} + 76\sigma_{o-e} +   
48\sigma_{o-o}}{344} =  0.9402
\end{equation}
From these results we may say that globally semFIS, with $\sigma_{semFIS}^B=
0.4269$, is the best one, followed in order by UNIV, $\sigma_{UNIV}^B =
0.5634$, ASAF, $\sigma_{newF}^B=0.6299$, and $\sigma_{ASAF}^B=0.6313$. 
Nevertheless, it is interesting to observe that for even-even alpha emitters
newF gives the 3rd best result after semFIS ($\sigma_{newF e-e}=0.298$
compared to $\sigma_{semFIS e-e}^B=0.223$ and $\sigma_{semFIS e-e}^C=0.287$), 
and for odd-even parents the 2nd
is newF with $\sigma_{newF o-e}^B=0.522$ compared to $\sigma_{semFIS
o-e}=0.442$.  newF is also better than ASAF for o-o nuclides, when
$\sigma_{newF o-o}^B=0.840$ and $\sigma_{ASAF o-o}^B=0.876$.

\section{Conclusions}

The accuracy of the new formula was increased after optimization of the five
parameters in the order: a; e; d; c, and b.  The SemFIS formula taking into
account the magic numbers of nucleons, the analytical super-asymmetric
fission model and the universal curves may be used to estimate the alpha
emitter half-lives in the region of superheavy nuclei.  The dependence on
the proton and neutron magic numbers of the semiempirical formula may be
exploited to obtain informations about the values of the magic numbers which
are not well known until now.

We introduced a weighted mean value of the rms standard deviation, allowing
to compare the global properties of a given model.  In this respect for the
set B the order of the four models is the following: semFIS; UNIV; newF, and
ASAF.  Nevertheless for even-even alpha emitters UNIV gives the 2nd best
result after semFIS, and for odd-even parents the 2nd is newF.

The quality of experimental data was also tested, as one can see by
comparing the three sets (A, B, C).  The set B with large number of emitters
(580) gives the best global result.  It is followed by the set A (454) three
times and the set C (344).

Despite its simplicity in comparison with semFIS the new formula, presented
in this article, behaves quite well, competing with the others well known
relationships discussed in the Ref. ~\cite{wan15prc}. 

\begin{acknowledgments} 

This work was supported within the IDEI Programme under Contracts No. 
43/05.10.2011 and 42/05.10.2011 with UEFISCDI, and NUCLEU Programme
PN16420101/2016 Bucharest.  

\end{acknowledgments}


\end{document}